\documentclass[twocolumn,twocolappendix]{aastex631}

\usepackage{graphicx}	
\usepackage{amsmath}	
\usepackage{multirow}
\usepackage{comment}
\usepackage{array}
\usepackage{soul}
\usepackage{etoolbox}
\usepackage{ae,aecompl}
\usepackage{hyperref}
\usepackage{threeparttable}
\usepackage{mathrsfs}
\usepackage{scalerel}
\usepackage{longtable,booktabs,threeparttablex}
\usepackage{float}

\definecolor{purple}{RGB}{160,32,240}
\definecolor{red}{RGB}{225,50,50}

\definecolor{addchange}{RGB}{215,25,25}

\definecolor{removechange}{RGB}{25,25,215}

\newcommand{\JWST}{\emph{JWST}}
\newcommand{\Spitzer}{\emph{Spitzer}}
\newcommand{\Muv}{\ensuremath{M_\mathrm{UV}^{ }}}

\newcommand{\OIIIHb}{[O\,{\sc iii}]+H\ensuremath{\beta}}
\newcommand{\OIII}{[O\,{\sc iii}]}
\newcommand{\OII}{[O\,{\sc ii}]}
\newcommand{\NeIII}{[Ne\,{\sc iii}]}

\newcommand{\CIIIsf}{C\,{\sc iii}]}
\newcommand{\CIIIf}{[C\,{\sc iii}]}
\newcommand{\CII}{[C\,{\sc ii}]}
\newcommand{\zCII}{\ensuremath{z_{_\mathrm{[CII]}}}}
\newcommand{\zOIII}{\ensuremath{z_{_\mathrm{[OIII]}}}}

\newcommand{\Msol}{\ensuremath{M_{\odot}}}

\newcommand{\logMstar}{\ensuremath{\log\left(M_{\ast}/M_{\odot}\right)}}

\newcolumntype{P}[1]{>{\centering\arraybackslash}p{#1}}

\defcitealias{RobertsBorsani2016}{RB16}

\shorttitle{Weak CIII] in UV-bright $z\sim7-8$ Galaxies}
\shortauthors{Endsley et al.}

\begin{document}

\title{REBELS-MOSFIRE: Weak CIII] Emission is Typical Among Extremely UV-bright, Massive Galaxies at $z\sim7$}

\author[0000-0003-4564-2771]{Ryan Endsley}
\affiliation{Department of Astronomy, University of Texas at Austin, 2515 Speedway, Austin, Texas 78712, USA}
\affiliation{Steward Observatory, University of Arizona, 933 N Cherry Avenue, Tucson, AZ 85721, USA}
\author[0000-0003-3509-4855]{Alice E. Shapley}
\affiliation{Department of Physics \& Astronomy, University of California, Los Angeles, 430 Portola Plaza, Los Angeles, CA 90095, USA}

\author[0000-0001-8426-1141]{Michael W. Topping}\affiliation{Steward Observatory, University of Arizona, 933 N Cherry Avenue, Tucson, AZ 85721, USA}
\author[0000-0001-6106-5172]{Daniel P. Stark}\affiliation{Department of Astronomy, University of California, Berkeley, Berkeley, CA 94720, USA}
\author[0000-0002-4989-2471]{Rychard J. Bouwens}
\affiliation{Leiden Observatory, Leiden University, NL-2300 RA Leiden, Netherlands}
\author[0009-0009-2671-4160]{Lucie E. Rowland}
\affiliation{Leiden Observatory, Leiden University, NL-2300 RA Leiden, Netherlands}
\author[0000-0002-2906-2200]{Laura Sommovigo}
\affiliation{Center for Computational Astrophysics, Flatiron Institute, New York, NY 10010, USA}
\author[0000-0002-4205-9567]{Hiddo S. B. Algera}
\affiliation{Institute of Astronomy and Astrophysics, Academia Sinica, 11F of Astronomy-Mathematics Building, No.1, Sec. 4, Roosevelt Rd, Taipeii 106319, Taiwan, R.O.C.}
\author[0000-0002-6290-3198]{Manuel Aravena}
\affiliation{Instituto de Estudios Astrof\'{\i}cos, Facultad de Ingenier\'{\i}a y Ciencias, Universidad Diego Portales, Av. Ej\'ercito 441, Santiago, Chile}
\affiliation{Millennium Nucleus for Galaxies (MINGAL)}
\author[0000-0003-3917-1678]{Rebecca A. A. Bowler}
\affiliation{Jodrell Bank Centre for Astrophysics, Department of Physics and Astronomy, School of Natural Sciences, The University of Manchester, Manchester M13 9PL, UK}
\author[0000-0001-9759-4797]{Elisabete da Cunha}
\affiliation{International Centre for Radio Astronomy Research, University of Western Australia, 35 Stirling Hwy, Crawley, WA 6009, Australia}
\affiliation{Research School of Astronomy and Astrophysics, Australian National University, Canberra, ACT 2611, Australia}
\affiliation{ARC Centre of Excellence for All Sky Astrophysics in 3 Dimensions (ASTRO 3D)}
\author[0000-0001-9419-6355]{Ilse de Looze}
\affiliation{Sterrenkundig Observatorium, Ghent University, Krijgslaan 281-S9, B-9000 Gent, Belgium}
\author[0000-0002-9400-7312]{Andrea Ferrara}
\affiliation{Scuola Normale Superiore, Piazza dei Cavalieri 7, 50126 Pisa, Italy}
\author[0000-0003-3917-1678]{Rebecca Fisher}
\affiliation{Jodrell Bank Centre for Astrophysics, Department of Physics and Astronomy, School of Natural Sciences, The University of Manchester, Manchester M13 9PL, UK}
\author[0000-0002-3120-0510]{Valentino Gonz\'alez}
\affiliation{Departamento de Astronom\'ia, Universidad de Chile, Camino del Observatorio 1515, Las Condes, Santiago 7591245, Chile}
\author[0000-0003-4268-0393]{Hanae Inami}
\affiliation{Hiroshima Astrophysical Science Center, Hiroshima University, 1-3-1 Kagamiyama, Higashi-Hiroshima, Hiroshima 739-8526, Japan}
\author[0000-0003-2804-0648]{Themiya Nanayakkara}
\affiliation{Centre for Astrophysics and Supercomputing, Swinburne University of Technology, P.O. Box 218, Hawthorn, 3122, VIC, Australia}
\author[0000-0001-9746-0924]{Sander Schouws}
\affiliation{Leiden Observatory, Leiden University, NL-2300 RA Leiden, Netherlands}
\author[0000-0001-5940-338X1]{Mengtao Tang}\affiliation{Steward Observatory, University of Arizona, 933 N Cherry Avenue, Tucson, AZ 85721, USA}

\correspondingauthor{Alice Shapley}
\email{aes@astro.ucla.edu}

\begin{abstract}

We present Keck/MOSFIRE $H$-band spectroscopic measurements covering the \CIIIf{}$\lambda 1907$, \CIIIsf{}$\lambda 1909$ doublet for a sample of 8 $z\sim 7$ spectroscopically-confirmed star-forming galaxies drawn from the Reionization Era Bright Emission Line Survey (REBELS). This REBELS-MOSFIRE sample is notable for its bright median UV luminosity ($\Muv{}=-22.5$~AB) and large median stellar mass ($\logMstar=9.2$). Although three sources show tentative evidence of a \CIIIsf{} detection, we obtain no confident detections for any of the 8 REBELS-MOSFIRE sources. The median \CIIIf{}$\lambda 1907$+\CIIIsf{}$\lambda 1909$ 3$\sigma$ upper limit in equivalent width (EW) for the REBELS-MOSFIRE sample is $6.5$\AA{}, and a stack of their $H$-band  MOSFIRE spectra yields a non-detection with an associated 3$\sigma$ upper limit of $2.6$\AA{}. These upper limits fall significantly below the \CIIIsf{} EW measured in a composite spectrum of representative $z\sim 7$ star-forming galaxies, as well as those measured for notable early star-forming galaxies such as GN-z11, GHZ2, GS-z12, and RXCJ2248-ID. The lack of strong \CIIIsf{} emission can be understood within the context of the stellar populations of the REBELS galaxies, as well as the ionization conditions and gas-phase metallicity implied by rest-frame optical spectroscopic properties (\OIIIHb{} EWs, and \OIII{}$\lambda5007$/\OII{}$\lambda3727$ and \NeIII{}$\lambda3869$/\OII{}$\lambda3727$ line ratios). The REBELS-MOSFIRE sample represents the higher-mass, higher-metallicity, lower-excitation tail of the $z\sim 7$ galaxy population, whose ionizing properties must be fully characterized to constrain the role of star-forming galaxies during cosmic reionization.

\end{abstract}


\section{Introduction} \label{sec:intro}
The spectroscopic frontier for galaxy formation now extends well into the first billion years of cosmic time \citep[e.g.,][]{Carniani2024_z14,Naidu2025}, bringing into focus the properties of galaxies during the epoch of reionization \citep{Stark2025}. A major outstanding goal is to understand the role played by star-forming galaxies in driving cosmic reionization. Spectroscopic probes of nebular emission lines provide powerful insights into the ionizing properties of star-forming galaxies. The strengths of emission lines in both the rest-frame UV and rest-frame optical constrain the nature of the ionizing radiation field, and the metallicity and chemical abundance pattern in the massive stars and gas \citep[e.g.,][]{Bunker2023_GNz11,Topping2024_RXCJ,Calabro2024}.

The first glimpse of these early ionizing properties was obtained prior to the launch of \JWST{}. \Spitzer{}/IRAC photometry was used to infer the mean and distribution of rest-optical (i.e., \OIIIHb{}) equivalent widths (EWs) for star-forming galaxies at $z\sim 7-8$ \citep{Labbe2013,Smit2014,Endsley2021_OIII,Stefanon2022_z8Ha}, demonstrating that these early galaxies were characterized by significantly stronger \OIIIHb{} emission on average than comparably massive galaxies observed at ``Cosmic Noon" (i.e., $z\sim 2$). This shift towards stronger rest-frame optical emission lines at earlier times was interpreted in terms of higher specific star-formation rates (sSFR) and lower gas-phase metallicities. At the same time, the MOSFIRE spectrograph \citep{McLean2012} on the Keck~I telescope was used to detect rest-frame UV emission lines from metals (i.e., \CIIIf{}$\lambda$1907, \CIIIsf{}$\lambda$1909 and CIV$\lambda\lambda 1548,1550$) for a small sample of luminous galaxies at $z\sim 7-8$ \citep[e.g.,][]{Stark2015,Stark2017,Laporte2017,Hutchison2019}. These detections reinforced a picture of low metallicity, accompanied by an intense and hard radiation field \citep{Topping2021}.

During the pre-\JWST{} era, it was challenging to assemble large samples of such rest-frame UV nebular metal emission-line measurements at $z\geq 7$ for many reasons. Perhaps most importantly, given the bright and highly-wavelength-dependent sky background affecting near-IR ground-based spectroscopic observations, a prior spectroscopic redshift was highly desirable for targeting \CIIIf{}$\lambda$1907, \CIIIsf{}$\lambda$1909 emission from the ground. Such spectroscopic redshifts at $z\geq 7$ were extremely small in number prior to the launch of \JWST{}. One of the largest pre-\JWST{} samples of spectroscopic redshifts came from the Reionization Era Bright Emission Line Survey \citep[REBELS][]{Bouwens2022_REBELS}. This ALMA large program targeted 40 luminous $\Muv{} < -21.5$ galaxies at $z\sim 6.5-9$, yielding a sample of 27 galaxies with \CII{}158$\mu$m detections (and therefore precise spectroscopic redshifts, Schouws et al. in prep.), and 16 with confirmed dust continuum emission \citep{Inami2022}. Therefore, the REBELS dataset from ALMA provides a unique and complementary window into the ISM for a well-defined sample of very distant galaxies.

Here we describe the results of a Keck/MOSFIRE program designed to detect \CIIIf{}$\lambda$1907, \CIIIsf{}$\lambda$1909 emission from a sample of 8 REBELS galaxies at $z\sim 7$ (hereafter referred to as the REBELS-MOSFIRE sample). At the time we conducted the observations (2021--2022), a unique strength of the REBELS targets was their confirmed spectroscopic redshifts, so critical for robust ground-based follow-up. Subsequent to our MOSFIRE observations, \JWST{} has transformed the field in terms of the sample of spectra for $z\geq 7$ galaxies. Yet, as we describe here, the REBELS-MOSFIRE sample occupies a relatively untested region of parameter space for rest-frame UV spectroscopy at $z\sim 7$, in terms of both its photometric \citep[e.g., stellar mass, rest-frame UV slope, dust continuum emission;][]{Bouwens2022_REBELS} and rest-frame optical spectroscopic properties \citep[e.g., \OIIIHb{} EW, \OIII{}$\lambda5007$/\OII{}$\lambda 3727$, and \NeIII{}$\lambda3869$/\OII{}$\lambda3727$;][]{Rowland2025}. Therefore, the measurements presented here represent a vital component of a complete description of the rest-UV spectroscopic properties of the $z\sim 7$ star-forming galaxy population.

In Section~\ref{sec:sec2}, we describe our REBELS-MOSFIRE sample and MOSFIRE observations. In Section~\ref{sec:results}, we present the key results of these MOSFIRE observations, both for individual objects and for our sample as a whole. Section~\ref{sec:discussion} considers the larger context and implications of these results, and we summarize in Section~\ref{sec:summary}.
Throughout this paper, we quote all magnitudes in the AB system \citep{Oke1983}, report EWs in rest-frame units, assume a \citet{Chabrier2003} stellar initial mass function (IMF) with limits of 0.1--300 \Msol{}, and adopt a flat $\Lambda$CDM cosmology with parameters $h=0.7$, $\Omega_\mathrm{M}=0.3$, and $\Omega_\mathrm{\Lambda}=0.7$. 

\section{Sample and Observations} \label{sec:sec2}

We first motivate and describe the target sample in \S\ref{sec:sample}, and next detail our Keck/MOSFIRE observations of this sample in \S\ref{sec:observations}.

\begin{figure}
\includegraphics[width=\columnwidth]{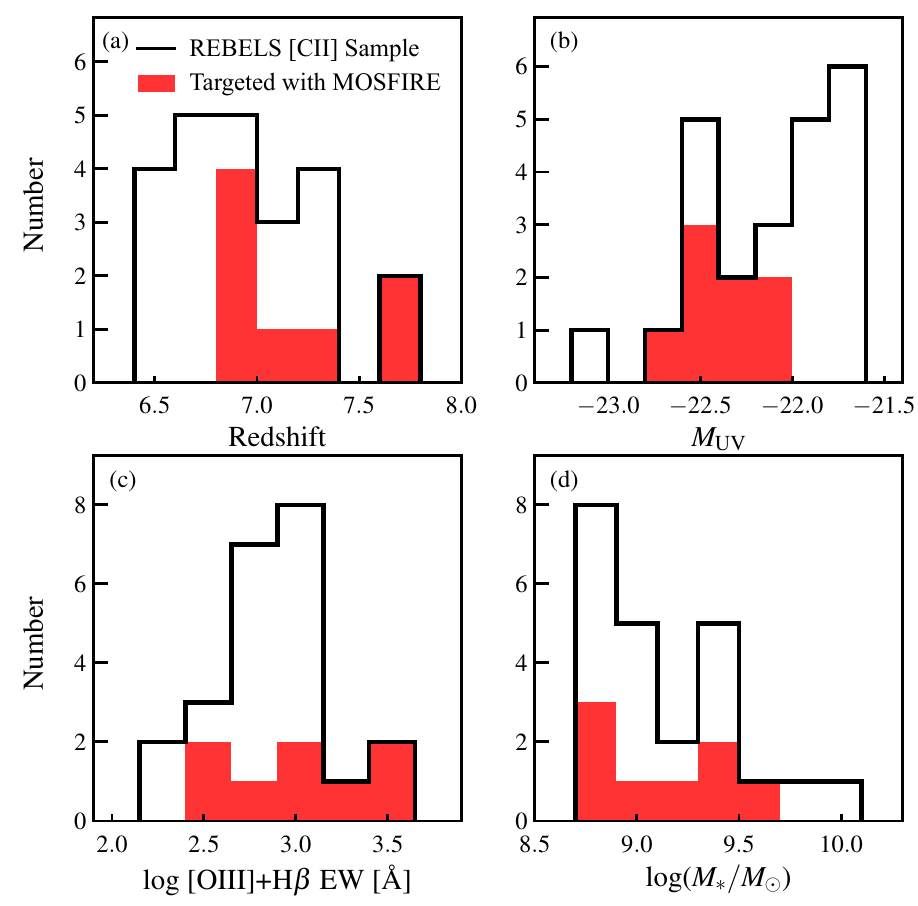}
\caption{The distributions of (a) spectroscopic redshifts, (b) absolute UV magnitudes, (c) photometric \OIIIHb{} EWs, and (d) photometric stellar masses for the parent sample of REBELS galaxies with ALMA \CII{} detections (black) and the REBELS-MOSFIRE subset targeted for \CIIIsf{} follow up (red).}
\label{fig:sampleDemographics}
\end{figure}

\subsection{Target Sample from REBELS} \label{sec:sample}

We aim to statistically characterize the strength of the nebular \CIIIf{}$\lambda$1907, \CIIIsf{}$\lambda$1909 doublet (hereafter shorthanded as ``\CIIIsf{} doublet") in reionization-era galaxies. 
For this purpose, we have chosen to target a subset of extremely UV-bright ($-23 \lesssim \Muv \lesssim -22$) $z\sim 7-8$ star-forming galaxies from the ALMA REBELS survey \citep{Bouwens2022_REBELS}.
There are two primary reasons why the REBELS galaxies are the optimal targets for this study.
First, their systemic \CII{}158$\mu$m redshift measurements precisely determine the observed wavelength of each \CIIIsf{} doublet component. This information is essential for ground-based near-IR spectroscopic follow-up observations, in which the measurement of \CIIIsf{} is performed against a bright and strongly wavelength-dependent sky background.
Second, the extremely luminous nature of the REBELS galaxies should make even moderate \CIIIsf{} emission (EW$\approx$3--5 \AA{}) accessible with ground-based facilities. 

For our \CIIIsf{} follow-up, we prioritized REBELS galaxies according to the following criteria: (1) there is a confident \CII{}158$\mu$m redshift measurement; (2)  the \CII{}158$\mu$m redshift places \CIIIsf{} in a region of high atmospheric and instrument throughput; and (3) at least one (preferably both) of the \CIIIsf{} doublet components is free of strong OH skyline contamination.
Among those with robust \CII{} redshifts, we also prioritized targeting a subset of REBELS galaxies that span a wide range of rest-optical \OIIIHb{} EWs given known correlations at lower redshift \citep[e.g.,][]{Tang2021}.
To estimate the \OIIIHb{} EW of each REBELS galaxy, we fit their 1--5$\mu$m photometry from VISTA/VIRCam and \Spitzer{}/IRAC with the BayEsian Analysis of GaLaxy sEds (\textsc{beagle}) code \citep{Chevallard2016}.
\textsc{beagle} fits against stellar photoionization models \citep{Gutkin2016} produced by running the updated \citet{Bruzual2003} stellar population synthesis models through the photoionization code, Cloudy (c13.03; \citealt{Ferland2013}).
We refer the interested reader to \citet{Endsley2022_REBELS} for details of the SED fitting procedure.
Briefly, we adopt a two-component `delayed+burst' star formation history along with a wide prior range in ionization parameter ($-4 \leq \mathrm{log}\,U \leq -1$) and metallicity ($-2.2 \leq \mathrm{log}(Z/Z_\odot) \leq 0.24$). This represents both the stellar and interstellar medium (ISM) metallicities in the \textsc{beagle} models, though the effective gas-phase metallicity is modulated by dust grain depletion via the dust-to-metal mass ratio, which we allow to vary in the range $\xi_d = 0.1-0.5$ (see \citealt{Gutkin2016}). We also fit the effective V-band dust attenuation to lie in the range $-3 \leq \log_{10}(\tau_V) \leq 0.7$ using the SMC attenuation law \citep{Pei1992}. As motivated in \citet{Endsley2021_OIII}, the SMC law well matches the IRX--$\beta$ relation at $z\sim2-3$ \citep{Bouwens2016, Reddy2018}.

The IDs, \CII{} redshifts, absolute UV magnitudes, as well as the photometrically-inferred \OIIIHb{} EWs and stellar masses of the REBELS-MOSFIRE galaxies are reported in Table~1. 
The redshifts span $\zCII{} = 6.845-7.677$ with an average value of $\langle \zCII{} \rangle = 7.2$. 
Because we prioritized the brightest REBELS galaxies (thus delivering the highest sensitivity to \CIIIsf{} emission), all of our targets have absolute UV magnitudes in the range $-22.7 \leq \Muv{} \leq -22.2$, which translates to 4.4--6.9$\times$ the characteristic UV luminosity at $z\sim7$ ($M_\mathrm{UV}^\ast = -20.6$; \citealt{Bowler2017}). 
The photometrically-inferred \OIIIHb{} EWs of the REBELS-MOSFIRE sample span $\lesssim$300 \AA{} to $\sim$3000--4000 \AA{}, thus fully encompassing the broad range of \OIIIHb{} EWs among UV-bright $z\sim7$ galaxies \citep{Endsley2021_OIII,Endsley2024_JADES,Begley2024,Meyer2024}.
As expected from their extremely UV luminous nature, all of the REBELS-MOSFIRE galaxies are inferred to be relatively massive with $8.8 \leq \logMstar{} \leq 9.6$ based on the SED fitting methodology described above. 
In Fig.~\ref{fig:sampleDemographics}, we show how the distributions of redshifts, absolute UV magnitudes, \OIIIHb{}, and stellar masses of the REBELS-MOSFIRE sample compare to those of the parent REBELS sample (including only those with ALMA \CII{} redshift measurements).

Recently, \citet{Rowland2025} reported results from \JWST{}/NIRSpec IFU observations of four of the eight REBELS-MOSFIRE galaxies. Hereafter, we refer to this sample of 12 REBELS galaxies with NIRSpec IFU observations as the ``REBELS-IFU'' sample.
In Table~1, we report the measured \OIIIHb{} EWs and inferred stellar masses from the IFU data as well.
Notably, we identify a systematic difference between the \OIIIHb{} EWs inferred from the ground-based near-infrared and IRAC photometry versus those measured from the IFU spectra.
The photometrically-inferred \OIIIHb{} EWs are all a factor of $\approx$2--3 higher than the spectroscopically-measured EWs.
Because higher EWs imply a more O-star dominated system with lower mass-to-light ratio, this in turn results in systematically lower inferred stellar masses when using the photometry.

These significant differences in EWs and stellar masses likely arise (at least in part) from different methods in how the integrated light of each system was calculated.
\citet{Rowland2025} use the integrated light of all detected clumps within a system, where these clumps are sometimes separated by several kpc or were too faint to detect in the ground-based near-IR or IRAC imaging.
In comparison, the photometric values were derived using circular apertures centered on the ground-based near-IR imaging centroid and hence may have missed extended light or entire clumps.

A detailed investigation of the origin of the systematic differences is beyond the scope of this paper \citep[see also][]{Lines2025} and any such conclusions would remain tentative given the small sample size.
Here, we simply note that the average stellar mass of the REBELS-MOSFIRE sample may be $\sim$2--3$\times$ higher than that implied by the photometric inferences once accounting for extended light as possible with the IFU analysis. We also note that \citet{Topping2022_REBELS} find that stellar masses of REBELS galaxies inferred assuming \textsc{Prospector} \citep{Johnson2021} non-parametric star-formation histories are systematically higher than those inferred from \textsc{beagle} constant star-formation histories. The median mass for the REBELS-MOSFIRE sample assuming non-parametric star-formation histories is \logMstar{}$_{\rm med}$=9.6. Both of these points only strengthen our conclusion that the REBELS-MOSFIRE sample traces the high-mass tail of the $z\sim7$ star-forming galaxy population (see Section~\ref{sec:discussion}). 

\movetableright=-1in
\begin{table*}
\centering
\begin{threeparttable}
\caption{Summary of the REBELS-MOSFIRE galaxy properties and resulting \CIIIsf{} constraints\tnote{$\dagger$}}
\begin{tabular}{crrrrrrrr}
\hline
REBELS ID & R-12 & R-15 & R-18 & R-27 & R-28 & R-30 & R-36 & R-39 \\[4pt]
\hline
\multicolumn{9}{c}{Galaxy Properties} \\[2pt]
\zCII{} & 7.346 & 6.875 & 7.675 & 7.090 & 6.943 & 6.982 & 7.677 & 6.845 \\[4pt]
\Muv{} & $-22.5^{+0.2}_{-0.2}$ & $-22.6^{+0.1}_{-0.1}$ & $-22.4^{+0.1}_{-0.1}$ & $-22.2^{+0.2}_{-0.1}$ & $-22.6^{+0.1}_{-0.1}$ & $-22.3^{+0.1}_{-0.1}$ & $-22.2^{+0.1}_{-0.1}$ & $-22.7^{+0.1}_{-0.1}$ \\[4pt]
$\beta$ & $-2.0^{+0.5}_{-0.8}$ & $-2.2^{+0.5}_{-0.5}$ & $-1.3^{+0.2}_{-0.3}$ & $-1.8^{+0.4}_{-0.4}$ & $-2.0^{+0.3}_{-0.3}$ & $-2.0^{+0.2}_{-0.2}$ & $-2.6^{+0.5}_{-0.5}$ & $-2.0^{+0.4}_{-0.4}$ \\[4pt]
EW$_{{\rm [OIII]+H}\beta}^\mathrm{\,phot}$ [\AA{}]\tnote{a} 
& $1810^{+930}_{-640}$ & $4370^{+2030}_{-1950}$ & $640^{+470}_{-290}$ & $340^{+360}_{-200}$ & $920^{+540}_{-380}$ & $400^{+760}_{-250}$ & $970^{+720}_{-380}$ & $3370^{+1150}_{-960}$ \\[4pt]
EW$_{{\rm [OIII]+H}\beta}^\mathrm{\,spec}$ [\AA{}]\tnote{b} 
& $650^{+35}_{-35}$ & $1826^{+42}_{-42}$ & $366^{+17}_{-17}$ & \nodata & \nodata & \nodata & \nodata & $1525^{+42}_{-42}$ \\[4pt]
Phot. \logMstar{}\tnote{a} & $9.1^{+0.4}_{-0.3}$ & $9.2^{+0.4}_{-0.3}$ & $9.3^{+0.4}_{-0.5}$ & $9.6^{+0.3}_{-1.0}$ & $8.8^{+0.4}_{-0.3}$ & $9.4^{+0.3}_{-0.5}$ & $8.8^{+0.5}_{-0.4}$ & $8.9^{+0.4}_{-0.2}$ \\[4pt]
Spec. \logMstar{}\tnote{b} & $9.54^{+0.03}_{-0.04}$ & $9.31^{+0.02}_{-0.01}$ & $9.71^{+0.06}_{-0.04}$ & \nodata & \nodata & \nodata & \nodata & $9.35^{+0.09}_{-0.08}$ \\[4pt]
\hline
\multicolumn{9}{c}{MOSFIRE Observations and Results} \\[2pt]
Exposure Time [hr] & 6.0 & 3.8 & 3.5 & 5.0 & 7.5 & 3.2 & 3.4 & 3.5 \\[4pt]
Seeing\tnote{c}\,\, [arcsec] & 0.57 & 0.70 & 0.73 & 0.55 & 0.62 & 0.60 & 0.59 & 0.63 \\[4pt]
PA [deg] & 8.0 & $-$79.0 & 161.5 & 200.0 & 77.0 & $-$16.0 & $-$40.0 & $-$52.0 \\[4pt]
\CIIIf{}$\lambda1907$ Flux\tnote{d} & \multirow{2}{*}{$<$1.5} & \multirow{2}{*}{$<$5.2} & \multirow{2}{*}{$<$1.7} & \multirow{2}{*}{$<$4.4} & \multirow{2}{*}{$<$1.5} & \multirow{2}{*}{$<$1.2} & $<$1.5 & \multirow{2}{*}{$<$1.5} \\[1pt]
[$10^{-18}$ erg/s/cm$^2$] &  &  &  &  &  &  & (1.8$\pm$0.6)\tnote{e} &  \\[4pt]
\CIIIf{}$\lambda1907$ EW\tnote{d} & \multirow{2}{*}{$<$2.7} & \multirow{2}{*}{$<$8.7} & \multirow{2}{*}{$<$3.2} & \multirow{2}{*}{$<$10.3} & \multirow{2}{*}{$<$3.0} & \multirow{2}{*}{$<$2.3} & $<$3.5 & \multirow{2}{*}{$<$2.4} \\[1pt]
[\AA{}] &  &  &  &  &  &  & (4.4$\pm$1.3)\tnote{e} &  \\[4pt]
\CIIIsf{}$\lambda$1909 Flux\tnote{d} & \multirow{2}{*}{$<$1.3} & $<$5.7 & \multirow{2}{*}{$<$3.7} & \multirow{2}{*}{$<$4.5} & \multirow{2}{*}{$<$1.3} & \multirow{2}{*}{$<$2.2} & \multirow{2}{*}{$<$1.2} & $<$1.3 \\[1pt]
[$10^{-18}$ erg/s/cm$^2$] &  & (6.1$\pm$1.8)\tnote{e} &  &  &  &  &  & (1.3$\pm$0.6)\tnote{e} \\[4pt]
\CIIIsf{}$\lambda1909$ EW\tnote{d} & \multirow{2}{*}{$<$2.4} & $<$9.5 & \multirow{2}{*}{$<$6.8} & \multirow{2}{*}{$<$10.4} & \multirow{2}{*}{$<$2.5} & \multirow{2}{*}{$<$4.3} & \multirow{2}{*}{$<$3.0} & $<$2.2 \\[1pt] 
[\AA{}] &  & (10.2$\pm$2.7)\tnote{e} &  &  &  &  &  & (2.1$\pm$0.6)\tnote{e} \\[4pt]
\hline
\end{tabular}
\begin{tablenotes}
\item[a] \OIIIHb{} EWs and stellar masses inferred using the ground-based near-IR and \Spitzer{}/IRAC photometry described in \citet{Bouwens2022_REBELS}.
\item[b] \OIIIHb{} EWs measured and stellar masses inferred using \JWST{}/NIRSpec IFU spectra where available \citep{Rowland2025}. 
\item[c] Measured from the stacked weighted spectra.
\item[d] For non-detections, we quote 3$\sigma$ upper limits.
\item[e] Measurements for the tentative ($\approx$3.5$\sigma$) detections of \CIIIsf{}$\lambda1909$ in REBELS-15 and REBELS-39, as well as \CIIIf{}$\lambda1907$ in REBELS-36.
\item[$\dagger$] \CIIIsf{} constraints reported here and based on MOSFIRE observations are consistent with those inferred from NIRSpec IFU data (Stefanon et al., in prep.). 
\end{tablenotes}
\end{threeparttable}
\label{tab:table1}
\end{table*}

\subsection{MOSFIRE Observations and Data Reduction} \label{sec:observations}

Our \CIIIsf{} follow-up observations were performed with the MOSFIRE instrument  on the 10 m Keck I telescope.
In general, we aimed to reach a 3$\sigma$ rest-frame EW limit of $\approx$3--5 \AA{} for each doublet component.
Past work \citep{Stark2017,Hutchison2019} has demonstrated that this EW limit requires approximately 3.5 hours of exposure time with MOSFIRE under good conditions given the UV magnitudes of our target galaxies.
Our observations were taken across four semesters (2021A, 2021B, 2022A, and 2022B), and through a mixture of weather conditions.
In total, we obtained $\approx$3.5 hours of exposure time under ideal conditions ($\sim$0.6\arcsec{} seeing and clear skies) for half of our targets.
Others (REBELS-12, REBELS-15, REBELS-27, and REBELS-28) were observed under variable observing conditions (at times patchy clouds, strong winds, or poor seeing), requiring longer total exposure times to reach our intended EW limit.

All observations were conducted in the $H$ band using a slit width of 0.7\arcsec{}, yielding a resolving power of $R\sim3660$.
This set-up resolves the \CIIIsf{} doublet and mitigates OH skyline contamination.
We adopted a two-position (``A-B") dither pattern with an amplitude of $\pm$1.5\arcsec{}, and an integration time of 120 seconds per exposure. 

The two-dimensional MOSFIRE data were reduced using the fully automated pipeline described in \citet{Kriek2015}. 
To summarize, this custom IDL pipeline implements flat fielding and sky subtraction, masks bad pixels and cosmic rays, rectifies and wavelength calibrates the spectra in individual exposures, combines data from all exposures (for a given slit mask), and performs telluric correction along with an initial flux calibration.
Slit stars were placed on each mask to measure precise relative offsets between individual exposures, determine the seeing and relative weight of each exposure, and perform a final absolute flux calibration for each mask.
Exposure weights were computed as the maximum flux of the best-fitting Gaussian to the slit star's 2D spectral profile (see \citealt{Kriek2015} for details), which accounts for both relative throughput (due to, e.g., sky transparency) and seeing.
The seeing values reported in Table~1 were derived by fitting the 2D spectral profiles of the slit stars in the weighted stacked spectra.

For absolute flux calibration, we multiplied the final stacked 2D spectra by the ratio of the slit star's VISTA/VIRCam \textit{H} band photometry to its synthetic photometry from the spectrum (using the median ratio for masks with multiple slit stars).
We ensured that all slit stars were isolated and unsaturated in the VISTA Deep Extragalactic Observations (VIDEO; \citealt{Jarvis2013}) or UltraVISTA \citep{McCracken2012} imaging, which we used to compute their photometry. 
To account for the likely extended morphology of our extremely UV-bright $z\sim7$ targets, we implemented additional slit loss corrections using the size - \Muv{} relation from \citet{CurtisLake2016} adopting a S\'{e}rsic index of $n = 1$.

We utilized optimal extraction \citep{Horne1986} to obtain 1D spectra for each object.
For cases where we did not detect a significant emission-line feature, the optimal extraction was centered on the expected spatial position of the object (as determined by the reduction pipeline) with a full-width at half maximum (FWHM) equivalent to the seeing measured from slit stars.
In cases of non-detections, we calculated upper limits on the line flux by integrating over 1.5$\times$ the FWHM along the spectral axis, centered on the wavelength expected from the \CII{} redshift.

\section{Results} \label{sec:results}

\begin{figure*}
\includegraphics[width=\textwidth]{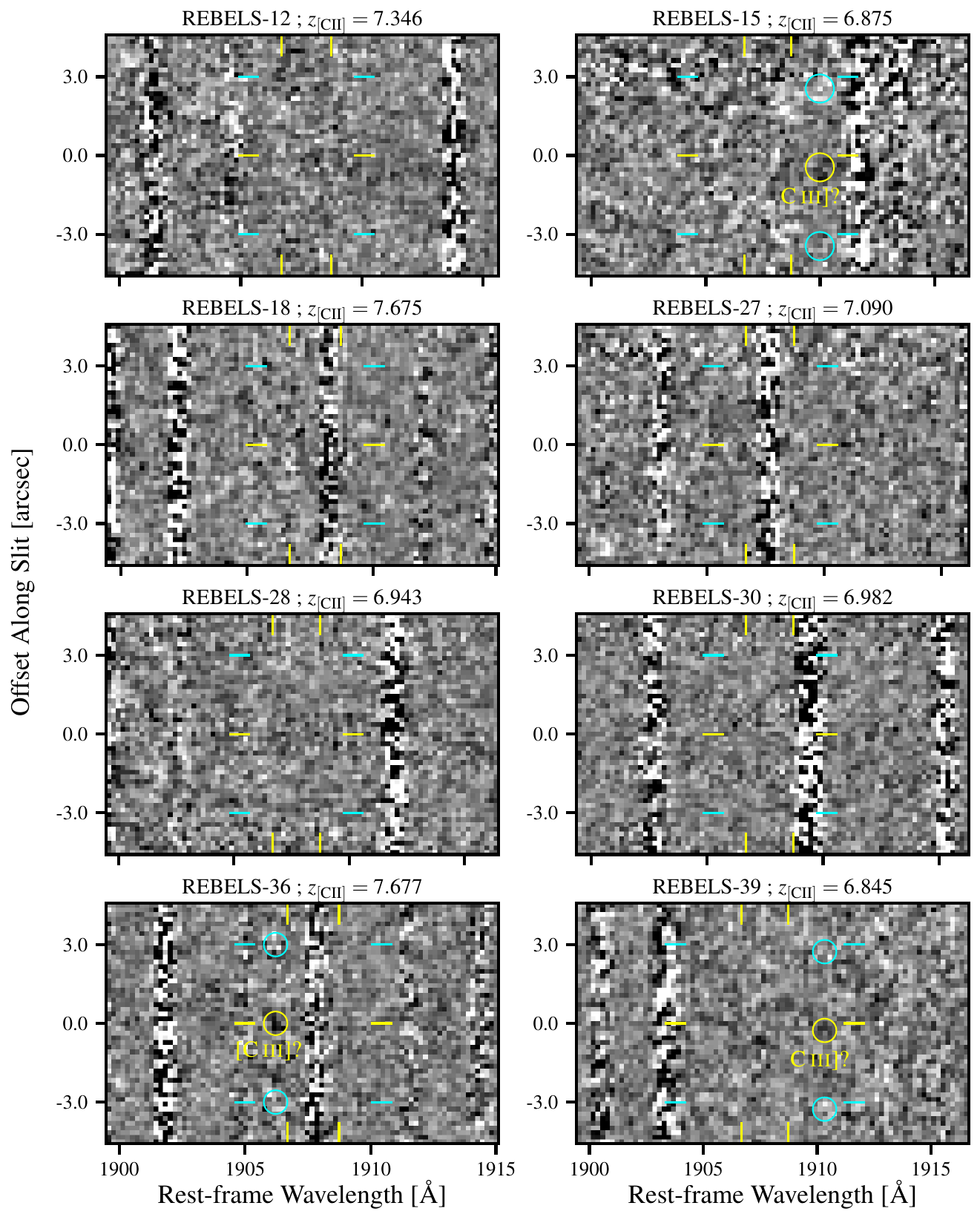}
\caption{The 2D MOSFIRE \textit{H}-band spectra around the \CIIIsf{} doublet in our eight targeted REBELS-MOSFIRE $z\sim7$ galaxies. Black is positive flux. Yellow marks indicate the expected wavelength and spatial positions of the \CIIIf{}$\lambda$1907 and \CIIIsf{}$\lambda$1909 features, while the cyan marks indicate where we would expect negative signal given the $\pm$1.5\arcsec{} A--B nodding dither pattern. We do not confidently detect any doublet feature. Tentative ($\approx$3.5$\sigma$) detections in REBELS-15, 36, and 39 are highlighted. The spectral regions with noticeably higher noise correspond to the wavelengths of known OH emission lines in the background sky spectrum, and reinforce the challenging nature of our ground-based \CIIIsf{} measurements due to OH skyline contamination.}
\label{fig:2Dspectra}
\end{figure*}

In this section, we detail the rest-UV \CIIIsf{}
constraints from our MOSFIRE observations of eight \CII{}-confirmed REBELS-MOSFIRE galaxies. 
Each REBELS-MOSFIRE target is discussed in turn, and the resulting flux and EW constraints for each doublet component are summarized in Table~1. The lack of strong \CIIIsf{} detections in the MOSFIRE spectra presented here is consistent with the \CIIIsf{} upper limits derived from the NIRSpec IFU spectra of REBELS galaxies (Stefanon et al., in prep.).

\subsection{REBELS-12} \label{sec:REBELS12}

REBELS-12 is an extremely UV-bright ($\Muv{} = -22.5\pm0.2$) galaxy with an ALMA \CII{}158$\mu$m redshift of \zCII{} = 7.346 and is detected in dust continuum \citep{Inami2022}. 
This object has a blue rest-UV slope ($F_\lambda \propto \lambda^{\beta}$) of $\beta = -2.0^{+0.5}_{-0.8}$ yet shows a strong excess in the IRAC 4.5$\mu$m band ([3.6]$-$[4.5] = $0.9\pm0.2$) implying a very high \OIIIHb{} EW of $1810^{+930}_{-640}$ \AA{}.
\citet{Rowland2025} report an \OIIIHb{} EW of 650$\pm$35 \AA{} using \JWST{}/NIRSpec IFU data.

We observed REBELS-12 with MOSFIRE under variable conditions on Dec. 15, 2021 ($t_\mathrm{exp}$ = 3.5 hours) and then again under good conditions Sep. 12, 2022 ($t_\mathrm{exp}$ = 2.5 hours), with the weighted stacked spectrum yielding 0.57\arcsec{} seeing.
No significant emission features are detected at the expected wavelengths of the \CIIIsf{} doublet in the stacked spectrum.
Given the ALMA redshift, neither component overlaps with a significant OH skyline feature and we accordingly achieve stringent 3$\sigma$ EW upper limits of 2.7 \AA{} and 2.4 \AA{} for \CIIIf{}$\lambda1907$ and \CIIIsf{}$\lambda1909$, respectively. \footnote{\citet{Algera2024} measure a redshift of \zOIII{}=7.3491 from \OIII{} 88$\mu$m emission. Along with the broad \CII{} emission profile for REBELS-12, this redshift difference is indicative of a possible merger scenario for REBELS-12. We do not detect \CIIIsf{} at a redshift of 7.3491 or 7.346.}

\subsection{REBELS-15} \label{sec:REBELS15}

REBELS-15 has $\Muv{} = -22.6\pm0.1$, \zCII{} = 6.845, and also shows a blue UV continuum ($\beta = -2.2\pm0.5$). It is one of four REBELS-MOSFIRE galaxies with published spectroscopic Ly$\alpha$ observations, and shows a rest-frame Ly$\alpha$ EW=3.7$\pm$0.8\~AA in emission \citep{Endsley2022_REBELS}. This galaxy is undetected in dust continuum \citep{Inami2022}, and shows a prominent IRAC excess ([3.6]$-$[4.5]=$-1.2\pm0.3$). 
At the \CII{} redshift, the \OIII{}5007 line lies at the far red edge of the [3.6] band with only 7\% max throughput, meaning that the IRAC color excess should be dominated by much weaker \OIII{}4959 and H$\beta$ emission.
Consequently, the strong [3.6]$-$[4.5] color implies an extremely high \OIIIHb{} EW of $4370^{+2020}_{-1950}$ \AA{}.
\citet{Rowland2025} measure a much lower (though still relatively extreme) EW of 1826$\pm$42 \AA{} from the NIRSpec IFU data.

MOSFIRE observations were conducted the night of Dec. 02, 2022 ($t_\mathrm{exp}$ = 3.8 hours) with relatively low transparency due to thin cirrus clouds, and we recover 0.70\arcsec{} seeing in the weighted stacked spectrum.
At the ALMA redshift, neither \CIIIsf{} component is contaminated by strong skylines and we do not identify any clear emission features.
We therefore adopt fiducial 3$\sigma$ upper EW limits of 8.7 \AA{} and 9.5 \AA{} for the \CIIIf{}$\lambda1907$ and \CIIIsf{}$\lambda1909$ components, respectively, for REBELS-15.

There is a tentative emission feature in the spectrum of REBELS-15 that may correspond to \CIIIsf{}$\lambda1909$ emission, though this requires additional data to verify.
This feature is located approximately 2.5 pixels ($\approx$0.45\arcsec{}) below the expected spatial location and is also offset from the expected wavelength (see Fig. \ref{fig:2Dspectra}).
After performing optimal 1D extraction at the spatial position of this feature, we measure a significance of 3.7$\sigma$ and a peak wavelength of $15041.6^{+1.6}_{-0.0}$ \AA{} translating to a redshift of $z=6.880^{+0.001}_{-0.000}$ if \CIIIsf{}$\lambda1909$. 
This is $\approx$200 km/s redward of that expected from the ALMA \CII{} systemic redshift.
Nonetheless, these wavelength and spatial offsets do not rule out the possibility that this feature is \CIIIsf{}$\lambda1909$ emission given the clumpy nature of REBELS-15.
As shown in Fig. 1 of \citet{Rowland2025}, recent \JWST{}/NIRSpec IFU data reveal that REBELS-15 is composed of at least two UV-bright clumps separated by $\approx$2 kpc ($\approx$0.4\arcsec{}) with a separation axis that approximately aligns with the position angle of our MOSFIRE observations.
It remains possible that the tentative feature at $\approx$15042 \AA{} is \CIIIsf{} emission arising from a star-forming clump with significant peculiar motions relative to the \CII{} line center. 
This clump may also be spatially offset from the blended centroid of rest-UV continuum emission in relatively poor angular resolution (FWHM$\approx$0.8\arcsec{}) ground-based imaging used to determine the expected spatial position of the \CIIIsf{} doublet.
If this feature is truly \CIIIsf{}$\lambda1909$ emission, it would have a line flux of (6.1$\pm$1.8)$\times$10$^{-18}$ erg/s/cm$^2$, corresponding to an EW of 10.2$\pm$2.7 \AA{}.

To better test the possibility that the feature is a real detection, we check whether negative signal is recovered in the weighted 2D spectrum at the expected spatial offsets given our dither pattern. 
After performing optimal extraction at these spatial positions and using the same wavelength integration window, 
we recover $-$4.5$\sigma$ and $-$0.8$\sigma$ signals at the two offset locations (see Fig. \ref{fig:2Dspectra}). 
The significance of the quadrature-summed negative signal is $-$4.6$\sigma$, which is consistent with the significance of the tentative positive \CIIIsf{}$\lambda1909$ emission feature.
Again, our fiducial conclusion is that the \CIIIsf{} doublet is undetected in REBELS-15, but stress that a tentative detection of fairly high EW \CIIIsf{}$\lambda1909$ emission awaits further assessment with deeper data.

\subsection{REBELS-18} \label{sec:REBELS18}

REBELS-18 exhibits a relatively red UV slope ($\beta = -1.3^{+0.2}_{-0.3}$) yet remains an extremely UV-bright ($\Muv{} = -22.4\pm0.1$) galaxy at \zCII{} = 7.675. It is also detected in dust continuum \citep{Inami2022}.
This system shows a moderate flux excess in the IRAC 4.5$\mu$m band ([3.6]$-$[4.5]=$0.4\pm0.2$) implying an \OIIIHb{} EW in the range $\approx$350--1100 \AA{}.
This is fairly consistent with the \OIIIHb{} EW of 366$\pm$17 \AA{} measured from NIRSpec IFU data \citep{Rowland2025}.

REBELS-18 was observed under mostly good conditions on Dec. 15, 2021 ($t_\mathrm{exp}$ = 3.5 hours) with the weighted stacked spectrum yielding 0.73\arcsec{} seeing.
We do not identify any significant emission features at the expected wavelengths of the \CIIIsf{} doublet, though we note that the \CIIIsf{}$\lambda1909$ component is expected to partially overlap with an OH skyline given the ALMA \CII{} redshift.
The data achieve 3$\sigma$ EW upper limits of 3.2 \AA{} and 6.8 \AA{} for the \CIIIf{}$\lambda1907$ and \CIIIsf{}$\lambda1909$ components, respectively.

\subsection{REBELS-27} \label{sec:REBELS27}

REBELS-27 has $\Muv{} = -22.2^{+0.2}_{-0.1}$, \zCII{} = 7.090, and a rest-UV slope of $\beta = -1.8 \pm 0.4$, typical of such bright $z\sim7$ galaxies \citep{Bowler2017,Endsley2021_OIII}. Spectroscopic Ly$\alpha$ observations of REBELS-27 show that it lacks Ly$\alpha$ emission, with an upper limit of $5.6-13.9$\AA{} in rest-frame EW \citep{Endsley2022_REBELS}. This galaxy is also detected in dust continuum \citep{Inami2022}.
We measure a flat IRAC color from REBELS-27 ([3.6]$-$[4.5]=$0.1\pm0.2$), implying a relatively low \OIIIHb{} EW of $340^{+360}_{-200}$ \AA{}.
REBELS-27 was observed for a total of 5.0 hours under partly cloudy conditions but with good seeing (0.55\arcsec{}) on Mar. 17, 2022, Apr. 9th, 2022, and Apr. 10th, 2022.
There are no significant emission features at the expected wavelengths of the \CIIIsf{} doublet, and both are free of strong OH skylines. 
Given the impact of low transparency, the data place relatively weak 3$\sigma$ EW upper limits of 10.3 \AA{} and 10.4 \AA{} for \CIIIf{}$\lambda1907$ and \CIIIsf{}$\lambda1909$, respectively.

\subsection{REBELS-28} \label{sec:REBELS28}

REBELS-28 lies at \zCII{} = 6.943, has $\Muv{} = -22.6\pm0.1$, and shows a blue UV continuum ($\beta = -2.0\pm0.3$). REBELS-28 is undetected in dust continuum \citep{Inami2022}, and exhibits a significantly blue IRAC color ([3.6]$-$[4.5]=$-1.4^{+0.4}_{-0.6}$) implying a fairly high \OIIIHb{} EW of $920^{+540}_{-380}$ \AA{}. 
We observed REBELS-28 on Mar. 28, 2021 ($t_\mathrm{exp}$ = 5.5 hours) with variable conditions, and again on Jan. 18, 2022 ($t_\mathrm{exp}$ = 2.0 hours) with good, stable conditions.
The weighted stacked spectrum yields 0.62\arcsec{} seeing.
We do not detect any significant emission feature at the expected wavelengths of the \CIIIsf{} doublet, and both doublet features are free of strong OH skylines.
The stacked data place stringent 3$\sigma$ EW upper limits of 3.0 \AA{} and 2.5 \AA{} for \CIIIf{}$\lambda1907$ and \CIIIsf{}$\lambda1909$, respectively.

\subsection{REBELS-30} \label{sec:REBELS30}

REBELS-30 has $\Muv{} = -22.3\pm0.1$, \zCII{} = 6.982, and exhibits a blue UV continuum ($\beta = -2.0\pm0.2$). REBELS-30 is undetected in dust continuum \citep{Inami2022}. Its marginally blue IRAC color ([3.6]$-$[4.5]=$-0.2^{+0.2}_{-0.2}$) implies an \OIIIHb{} EW of $400^{+750}_{-250}$ \AA{}. 
This system was observed under clear and stable conditions on Jan. 18, 2022 ($t_\mathrm{exp}$ = 3.2 hours), yielding 0.60\arcsec{} seeing in the weighted stacked spectrum.
No significant emission features are found at the expected wavelengths of the \CIIIsf{} doublet, though the \CIIIsf{}$\lambda1909$ component partially overlaps with a moderately strong OH skyline.
Accordingly, the data place 3$\sigma$ EW upper limits of 2.3 \AA{} and 4.3 \AA{} for \CIIIf{}$\lambda1907$ and \CIIIsf{}$\lambda1909$, respectively.

\subsection{REBELS-36} \label{sec:REBELS36}

REBELS-36 has $\Muv{} = -22.2\pm0.1$, \zCII{} = 7.677, and is undetected in dust continuum \citep{Inami2022}. This galaxy shows a very blue UV slope ($\beta = -2.6\pm0.5$) yet a significantly red IRAC color ([3.6]$-$[4.5]=$0.5^{+0.1}_{-0.1}$) suggesting a high \OIIIHb{} EW of $970^{+720}_{-320}$ \AA{}. It also shows Ly$\alpha$ in emission, with an EW of $24^{+5}_{-6}$\AA{} \citep{Valentino2022}. We observed this galaxy with MOSFIRE on Mar. 2, 2021 ($t_\mathrm{exp}$ = 3.4 hours) under good, stable conditions, and measure seeing of 0.59\arcsec{} in the weighted stack.
Neither \CIIIsf{} component is contaminated by OH skylines at the expected wavelength from the ALMA \CII{} redshift measurement, and we do not identify any obvious emission line features at either position. 
Accordingly, we adopt fiducial 3$\sigma$ upper EW limits of 3.5 \AA{} and 3.0 \AA{} for the \CIIIf{}$\lambda1907$ and \CIIIsf{}$\lambda1909$ components, respectively.

However, we do identify a tentative (3.4$\sigma$) emission line feature very near the expected spatial and wavelength position of the \CIIIf{}$\lambda1907$ line in REBELS-36 (see Fig. \ref{fig:2Dspectra}).
After performing optimal 1D extraction at the expected spatial position, we identify a feature at $16540.0^{+3.3}_{-0.0}$ \AA{}.
If \CIIIf{}$\lambda1907$, this would translate to a redshift of $z = 7.675^{+0.002}_{-0.000}$ consistent with the systemic ALMA redshift of \zCII{} = 7.677 within uncertainties.
From the optimal 1D extraction, we recover a line flux of (1.8$\pm$0.6)$\times$10$^{-18}$ erg/s/cm$^2$, corresponding to an EW of 4.4$\pm$1.3 \AA{}. 
We recover marginal negative signal when performing optimal extraction at the expected spatial offsets given our dither pattern ($-$1.5$\sigma$ below and $-$0.9$\sigma$ above), lending support that this may be a real feature.
Nonetheless, given that the significance of the feature itself is barely above 3$\sigma$, we consider this a tentative detection of \CIIIf{}$\lambda1907$ emission that requires further data to validate. 

\subsection{REBELS-39} \label{sec:REBELS39}

REBELS-39 lies at \zCII{} = 6.845, is extremely UV luminous ($\Muv{} = -22.7\pm0.1$), and exhibits both a flat UV continuum slope ($\beta = -2.0\pm0.4$) and Ly$\alpha$ in emission \citep[EW$=10.0\pm1.7$\AA{};][]{Endsley2022_REBELS}. This galaxy is also detected in dust continuum \citep{Inami2022}, and
shows a very strong excess in the 3.6$\mu$m IRAC filter ([3.6]$-$[4.5]=$-1.3^{+0.3}_{-0.3}$).
Because the \OIII{}5007 line is at only 24\% maximum transmission through the [3.6] band at the \CII{} redshift, the excess in [3.6] should be primarily driven by the weaker \OIII{}4959 and H$\beta$ lines.
This yields an extremely high photometrically-inferred \OIIIHb{} EW of $3370^{+1150}_{-960}$ \AA{}.
\citet{Rowland2025} report a much lower EW of 1525$\pm$42 \AA{} for REBELS-39, though their calculation incorporates light from another clump located $\approx$5 kpc (i.e., $\approx$1\arcsec{}) away from the clump on which the photometric apertures (and the MOSFIRE slit) are centered.

We observed REBELS-39 with MOSFIRE on Mar. 2, 2021 ($t_\mathrm{exp}$ = 3.5 hours) under good, stable conditions that resulted in 0.63\arcsec{} seeing in the weighted stacked spectrum.
Neither of the \CIIIsf{} doublet components overlap with an OH skyline at the wavelengths predicted from the ALMA \CII{} redshift, and we do not see any significant emission features at these positions.
We therefore adopt fiducial 3$\sigma$ upper EW limits of 2.4 \AA{} and 2.2 \AA{} for \CIIIf{}$\lambda1907$ and \CIIIsf{}$\lambda1909$, respectively.

However, similar to REBELS-15, we do identify a tentative emission feature in the spectrum of REBELS-39 that may correspond to \CIIIsf{}$\lambda1909$ emission.
This feature has a significance of 3.3$\sigma$ from the optimal 1D extraction, is located approximately 2 pixels ($\approx$0.35\arcsec{}) below the expected spatial location, and also has a peak wavelength of $14984.7^{+3.3}_{-0.0}$ \AA{}.
At the $\pm3$\arcsec{} dither offset locations we recover negative signal (see Fig. \ref{fig:2Dspectra}) with $-$3.3$\sigma$ and $-$1.9$\sigma$ significance summing to $-$3.8$\sigma$ in quadrature.
If \CIIIsf{}$\lambda1909$, the peak wavelength corresponds to a redshift of $z=6.851^{+0.002}_{-0.000}$ which is $\approx$230 km/s redward of the systemic redshift obtained from the ALMA \CII{} line center.
Nevertheless, as with REBELS-15, recent NIRSpec IFU data shows that REBELS-39 is also a clumpy system (see Fig. 1 of \citealt{Rowland2025}), so it is possible that this tentative feature really is \CIIIsf{}$\lambda1909$ emission arising from a star-forming clump with significant peculiar motions and spatially offset from the ground-based imaging centroid. 
If indeed \CIIIsf{}$\lambda1909$ emission, it would have a line flux of (1.3$\pm$0.6)$\times$10$^{-18}$ erg/s/cm$^2$, corresponding to an EW of 2.1$\pm$0.6 \AA{}.
Additional data is required to verify this tentative emission feature.

\subsection{REBELS-MOSFIRE Stack} \label{sec:stack}

Given the lack of any confident \CIIIsf{} detection among the individual galaxies in our REBELS-MOSFIRE sample, we now stack our MOSFIRE spectra to obtain more stringent constraints.
We first shift all spectra to the same rest-frame wavelength and spatial offset grid using the ALMA \CII{} systemic redshift and the ground-based rest-UV imaging centroid, then stack all eight 2D spectra using an inverse-variance weighting scheme.
The continuum flux densities of each source as well as the seeing from each MOSFIRE mask are stacked with the same weights.
We then perform optimal extraction at the expected spatial and wavelength positions of the expected doublets following the procedure outlined in \S\ref{sec:observations}.
Given the clumpy nature of many REBELS galaxies \citep{Rowland2025}, it is possible that not all will emit \CIIIsf{} photons at the expected spatial and rest-frame wavelength positions (see \S\ref{sec:REBELS15} and \S\ref{sec:REBELS39} for tentative examples). However, in the absence of additional information, this remains a systematic uncertainty in constructing the stacked \CIIIsf{} spectrum.

No significant emission feature is detected in the inverse-variance weighted REBELS-MOSFIRE stack. 
After integrating over 2$\times$ the FWHM along the spectral axis, we recover 3$\sigma$ flux upper limits of $<$0.7$\times$10$^{-18}$ erg/s/cm$^2$ for each doublet component.
This translates to a total \CIIIf{}$\lambda$1907 + \CIIIsf{}$\lambda$1909 EW 3$\sigma$ upper limit of $<$2.6 \AA{} when adopting the mean redshift of the REBELS-MOSFIRE sample.

\section{Discussion} \label{sec:discussion}

\begin{figure*}
\includegraphics[width=\textwidth]{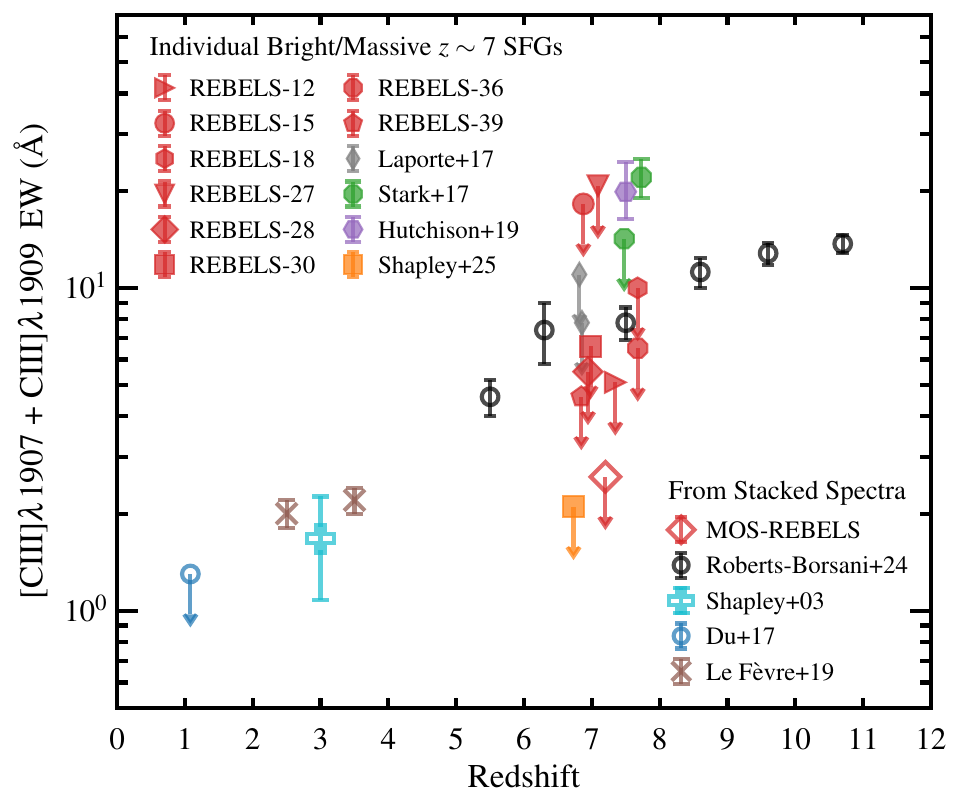}
\caption{Compilation of measurements of \CIIIsf{} EW over a broad range of redshifts. The majority of the REBELS-MOSFIRE sample (5 out of 8) have \CIIIsf{} EW upper limits (red markers) that fall below the average trend from \citet{RobertsBorsani2024} (black symbols). The dusty metal-rich $z=6.73$ galaxy GOODSN-100182 \citep[orange marker;][]{Shapley2025} also has a much lower \CIIIsf{} EW compared to typical $z\sim7$ galaxies \citep{Shapley2025}. These upper limits from REBELS-MOSFIRE and GOODSN-100182 highlight the diversity of \CIIIsf{} EWs among massive, luminous $z\sim 7$ galaxies relative to early work \citep{Stark2017,Hutchison2019}. The typical \CIIIsf{} EW of the REBELS-MOSFIRE galaxies appears similar to that of comparably massive ($M_\ast \sim 10^{9-10}\,\Msol{}$) galaxies at $z\sim1-4$ compiled by \citet{Du2017}, \citet{Shapley2003}, and \citet{LeFevre2019}.}
\label{fig:CIIIevolution}
\end{figure*}

Our deep Keck/MOSFIRE spectroscopy of 8 luminous ($\Muv{} \sim -22.5$), massive ($\logMstar{} \sim 9.0-9.5$) star-forming galaxies at $z=7-8$ has yielded no ironclad detections of \CIIIsf{} emission. The median 3$\sigma$ upper limit in \CIIIsf{} EW for our sample is $\sim 6.5$\AA{}. We emphasize that precise ALMA \CII{}158$\mu$m redshifts for our target galaxies \citep{Bouwens2022_REBELS} lead to tight constraints on the observed wavelengths at which we searched for \CIIIsf{} in the MOSFIRE spectra. We used these pre-existing \CII{}158$\mu$m redshifts to prioritize galaxies where \CIIIsf{} would appear relatively free of strong near-IR sky emission lines, enabling sensitive upper limits on the strength of \CIIIsf{} in the absence of robust detection.

Over the past two years, deep \JWST{}/NIRSpec data have confirmed that strong \CIIIsf{} emission (EW$\gtrsim$10 \AA{}) is common at $z>6$ \citep[e.g.,][]{Tang2023_CEERS,Stark2025}, supporting findings from early ground-based efforts \citep{Stark2015_CIII,Stark2017,Laporte2017,Mainali2018,Hutchison2019,Topping2021}. The evolution in typical \CIIIsf{} EW has been quantified by \citet{RobertsBorsani2024}, who stacked \JWST{}/NIRSpec prism spectra of 482 star-forming galaxies at $z = 5.0‑12.9$ in bins of redshift. This analysis included a stack of 54 galaxies at $z=7-8$ ($z_{\rm med} =7.47$), in which the  \CIIIsf{} EW is measured as $7.8\pm 0.9$\AA{}. As shown in Figure~\ref{fig:CIIIevolution}, the upper limits for 5 of the REBELS-MOSFIRE galaxies place them below the stacked average \CIIIsf{} EW from \citet{RobertsBorsani2024}. Looser \CIIIsf{} EW constraints for the remaining three REBELS-MOSFIRE galaxies are simply due to suboptimal observing conditions or, in one case, greater systematic uncertainty due to the coincidence of the predicted wavelength of the \CIIIsf{}$\lambda1909$ feature with a skyline. Overall, we infer that the median \CIIIsf{} EW for the REBELS-MOSFIRE sample must be significantly lower than what is reported by \citet{RobertsBorsani2024} for $z=7-8$, and similar to that of comparably massive star-forming galaxies at $z\sim 1-4$ \citep[i.e., $M_\ast \sim 10^{9-10}\,\Msol{}$;][]{Du2017,Shapley2003,LeFevre2019}.

\begin{figure*}
\includegraphics[width=\textwidth]{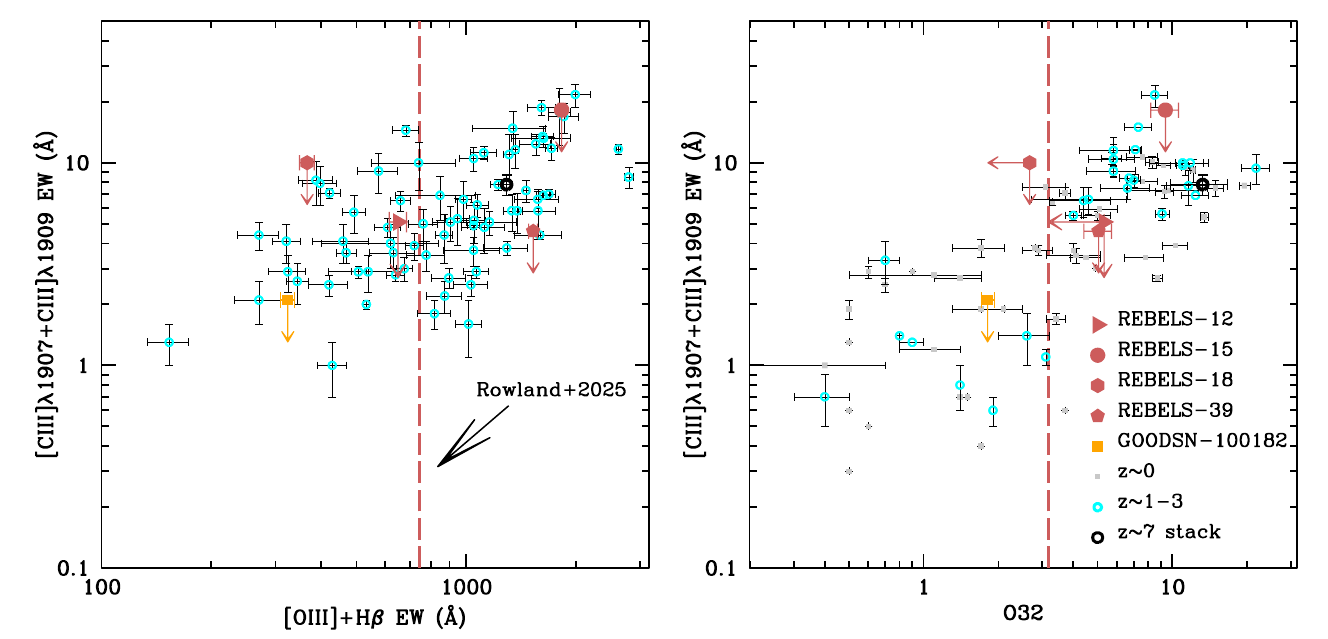}
\caption{\CIIIsf{} EW vs. rest-frame optical emission-line properties. Individual REBELS-MOSFIRE galaxies with NIRSpec rest-optical spectroscopy are plotted in red, while the median rest-frame optical properties of REBELS-IFU galaxies from \citet{Rowland2025} are indicated with dashed vertical lines. The massive, metal-rich $z=6.73$ galaxy GOODSN-100182 \citep{Shapley2025} is indicated with an orange square. Lower-redshift data compiled in \citet{Tang2021} and \citet{Tang2023_CEERS} are indicated in grey ($z\sim 0$) and cyan ($z\sim 1-3$). The $z\sim 7$ stacked values from \citet{RobertsBorsani2024} are shown in black. {\bf Left:} \CIIIsf{} EW vs. \OIIIHb{} EW. {\bf Right:} \CIIIsf{} EW vs. O32. In this panel, the vertical dashed line represents an upper limit on the REBELS-IFU sample median O32 from \citet{Rowland2025} (see Section~\ref{sec:discussion}).} 
\label{fig:c3restopt}
\end{figure*}

To understand this difference, we consider the nature of the REBELS-MOSFIRE galaxies in terms of both their stellar populations and ISM properties, in comparison with the median properties of the \citet{RobertsBorsani2024} sample. First, the REBELS-MOSFIRE sample is characterized by a median stellar mass of $\logMstar{}_{\rm med}=9.15$ (which is likely an underestimate; see \S\ref{sec:sample}), a median UV luminosity of $\Muv{}_{\rm med}=-22.5$, and a median UV slope of $\beta_{\rm med}=-2.0$. In contrast, the $z\sim 7$ stacked sample from \citet{RobertsBorsani2024} has $\logMstar{}_{\rm med}=7.89$, $\Muv{}_{\rm med}=-19.03$, and $\beta_{\rm med}=-2.5$. In \citet{RobertsBorsani2024}, $\beta$ values are determined by fitting a power-law model directly to NIRSpec prism spectra, whereas the values quoted here for the REBELS-MOSFIRE sample are inferred from the SED fitting described in Section~\ref{sec:observations}. However, \citet{Fisher2025} derive $\beta$ spectroscopically for the four REBELS-MOSFIRE galaxies with NIRSpec IFU data and find results that are consistent with our SED-fitting method.  Therefore, we can robustly state that the REBELS-MOSFIRE sample is, on average, more than an order of magnitude more massive and UV-luminous, and significantly redder in UV slope than the stacked $z\sim 7$ sample. We also note that 4 out of 8 of the REBELS-MOSFIRE sample have ALMA dust continuum 
detections, suggesting significant dust reservoirs \citep{Inami2022}. 

Second, the rest-optical emission-line properties of the REBELS-MOSFIRE sample are distinct from the average trends at $z\sim 7$. The median rest-frame \OIIIHb{} EW is 730\AA{} for the 4 galaxies in our sample with NIRSpec coverage, which is almost identical to the median of 745\AA{} for the larger sample of 12 REBELS-IFU galaxies presented by \citet{Rowland2025}. This median \OIIIHb{} EW for REBELS-MOSFIRE galaxies is significantly lower than that of the $z\sim 7$ stack from \citet{RobertsBorsani2024}, i.e., $1290\pm28$ \AA{}. Beyond \OIIIHb{} EW, we can consider rest-optical emission-line ratios that are sensitive to the excitation properties of the ionized ISM (e.g., the ionization parameter). 
These include \OIII5007/H$\beta$ (O3), \OIII5007/\OII3727 (O32) and \NeIII3869/\OII3727 (Ne3O2). Among the 4 REBELS-MOSFIRE galaxies with NIRSpec coverage, the median ratios are O3$_{\rm med}=6.9$, O32$_{\rm med}\leq5.18$\footnote{The median O32 value is listed here as an upper limit. REBELS-12 and REBELS-18 are both at $z>7$, such that H$\alpha$ falls redwards of the NIRSpec IFU prism wavelength coverage. \citet{Rowland2025} therefore present the dust-uncorrected O32 values as upper limits for these two objects. The combination of detected O32 values for REBELS-15 and REBELS-39 and O32 upper limits for REBELS-12 and REBELS-18 yields the upper limit quoted for the median. For the same reason, the median O32 for the full REBELS-IFU sample presented in \citet{Rowland2025} is also listed as an upper limit.} and Ne3O2$_{\rm med}=0.43$. As these 4 REBELS-MOSFIRE galaxies are representative of the REBELS-IFU sample presented in \citet{Rowland2025} in terms of stellar mass, $M_{{\rm UV}}$, rest-UV slope ($\beta$), and rest-frame \OIIIHb{} EW, we include the medians of this larger REBELS-IFU sample for better statistics (O3$_{\rm med}=5.5$, ranging from 4.0 to 7.5; O32$_{\rm med}\leq 3.18$, ranging from 1.88 to 9.38\footnote{The median O32 value for the REBELS-IFU sample is listed as an upper limit, because some of the individual O32 values reported in Table~4 of \citet{Rowland2025} are upper limits.}; and Ne3O2$_{\rm med}=0.35$, ranging from 0.23 to 0.84).  In the $z\sim 7$ stack from \citet{RobertsBorsani2024}, the rest-optical line ratios are O3=$5.3\pm0.3$, O32=$13.25\pm3.40$, and Ne3O2=$1.75\pm0.50$. These high O32 and Ne3O2 values are emblematic of all of the median stacks at $z=6-9$, accompanying uniformly blue values of UV slope (i.e., $\beta\leq -2.5$), stellar masses of $\logMstar{}\leq 8.0$, and UV luminosities fainter than $\Muv{}=-19.5$. Accordingly, there is no significant difference in O3, but the O32 and Ne3O2 ratios are significantly lower for the REBELS-IFU sample. 

In previous work, \citet{Tang2021} and \citet{Tang2023_CEERS} have used a large sample of star-forming galaxies at $z\sim 1-3$ to demonstrate the connection between rest-optical nebular emission-line properties and the strength of rest-UV emission lines such as \CIIIsf{}. In particular, significant positive correlations are observed between \CIIIsf{} EW and \OIIIHb{} EW, O32, and Ne3O2. As shown in Figure~\ref{fig:c3restopt}, based on the rest-frame optical properties of the REBELS-MOSFIRE galaxies presented here, the rest-frame optical properties of the larger REBELS-IFU sample presented by \citet{Rowland2025}, and the trends observed by \citeauthor{Tang2021} (e.g., the linear regression of \CIIIsf{} EW on O32), the typical predicted \CIIIsf{} EW for the REBELS-MOSFIRE sample is $\leq 5$\AA{}. Furthermore, given the lower \OIIIHb{} EW, O32 and Ne3O2 of the REBELS-IFU sample compared with the $z\sim 7$ stack from \citet{RobertsBorsani2024}, it is also predicted that the average \CIIIsf{} EW will be lower. We note that one of the REBELS-MOSFIRE galaxies with a tentative \CIIIsf{} detection, REBELS-15, is also the galaxy with \OIIIHb{} EW=$1826\pm42$\AA{}, O32=$9.38\pm 1.20$, and Ne3O2=$0.84\pm 0.24$. For each property, REBELS-15 has the highest value in the REBELS-IFU sample of \citet{Rowland2025} and among the subset of REBELS-MOSFIRE galaxies presented here. 

In physical terms, O32 and Ne3O2 are most directly measures of the ionization parameter \citep{Kewley2002,Levesque2014,Sanders2016}. However, given the anti-correlation between ionization parameter and gas-phase oxygen abundance observed in star-forming regions \citep{PerezMontero2014}, both line ratios can serve as indirect indications of nebular metallicity \citep[e.g.,][]{Sanders2024}. In particular, O32 and Ne3O2 are useful metallicity indicators in the event that the well-known nebular metallicity indicator, R23 (defined as ([OIII]5007,4959+[OII]3727)/H$\beta$), falls outside the range of strong-line calibrations. R23 (which requires a dust correction) is only measured for two of the four REBELS-MOSFIRE targets, REBELS-15 and REBELS-39, and for both of them it falls outside the calibrated range such that \citet{Rowland2025} do not use it as a metallicity indicator.

The average O32 and Ne3O2 
values for the REBELS-IFU sample suggest higher gas-phase oxygen abundances than represented by the $z\sim 7$ stack from \citet{RobertsBorsani2024}. Based on recent metallicity calibrations from large samples of {\it JWST}/NIRSpec faint auroral-line measurements of high-redshift star-forming galaxies \citep{Sanders2025}, the typical O32 and Ne3O2 values for the  REBELS-IFU sample correspond to metallicities of $\sim 0.3$~solar, where those from the $z\sim 7$ stack of \citet{RobertsBorsani2024} correspond to metallicities of $\sim 0.07$~solar. Using previously available metallicity calibrations, \citet{Rowland2025} estimated typical metallicities of $\sim 0.4$~solar for the sample of 12 REBELS-IFU galaxies with rest-frame optical strong-line measurements, whereas \citet{RobertsBorsani2024} estimate a metallicity of $\sim 0.1$~solar for their $z\sim 7$ stack. 
Regardless of the calibrations used, the same qualitative offset in metallicity is inferred between the REBELS-IFU sample and the $z\sim 7$ stack from \citet{RobertsBorsani2024}. This higher metallicity of the REBELS-IFU sample mirrors its difference in stellar mass, given the existence of the mass-metallicity relation at high redshift \citep{Nakajima2023,Curti2024}. Finally, many works \citep[e.g.,][]{Jaskot2016,Tang2021} have used photoionization modeling to highlight the strong connection between \CIIIsf{} EW and metallicity. At metallicities above $\sim 0.1$~solar, there is a strong decline in the strength of \CIIIsf{} as metallicity increases, reflecting the cooler nebular temperature, lower ionization parameter, and softer ionizing spectrum. The higher average metallicity of the REBELS-IFU sample provides a clear explanation for its weaker \CIIIsf{} than that of the $z\sim 7$ stack of \citet{RobertsBorsani2024}.

The FIR \OIII{}$88\mu$m/\CII{}$158\mu$m line ratio can also be used as a probe of the excitation conditions of the ionized ISM. In a pilot study of the FIR line ratios of two REBELS galaxies (REBELS-12, which is in the REBELS-MOSFIRE sample, and REBELS-25 at \zCII{}=7.3065), \citet{Algera2024} found significantly lower \OIII{}$88\mu$m/\CII{}$158\mu$m values compared with the range observed in other $z=6-9$ star-forming galaxies \citep{Harikane2020}. This difference provides further evidence of lower ionization parameters and/or more evolved stellar populations in REBELS galaxies compared to other star-forming galaxy samples at $z>6$.

The galaxy GOODSN-100182 represents perhaps an even more extreme version of the REBELS sample. As described in \citet{Shapley2025}, GOODSN-100182 ($z=6.73$) is characterized by $\logMstar{}=9.97^{+0.18}_{-0.24}$ and $\beta=-0.50\pm 0.09$. Therefore, it is both more massive and dustier than any object in the REBELS sample. GOODSN-100182 also appears to be metal-rich, with \OIIIHb{} EW=$324\pm14$\AA{}, O32=$1.81\pm 0.11$, and Ne3O2=$0.10\pm 0.02$. The NIRSpec G140M spectrum of GOODSN-100182 covers the wavelength of \CIIIsf{}, but yields only a significant detection of the continuum, with no line emission. As shown in Figures~\ref{fig:CIIIevolution} and \ref{fig:c3restopt}, the upper limit in \CIIIsf{} EW is 2.1~\AA{}. The rest-optical emission-line ratios of GOODSN-100182 are fairly typical for $z\sim 2-3$ main-sequence star-forming galaxies of comparable mass, which are characterized by average \CIIIsf{} EWs $\leq 2$\AA{} \citep{Steidel2016,Shapley2003}. Therefore, the REBELS sample and GOODSN-100182 represent the massive, metal-rich tail of the $z\sim 7$ star-forming galaxy population, in which \CIIIsf{} is typically weaker than 5\AA{} in rest-frame EW.

Such objects stand in stark contrast with notable examples of strong rest-frame UV nebular line emission in the very early universe. For example, among the highest-redshift galaxies with rest-frame UV emission-line detections, GN-z11 \citep[$z=10.60$;][]{Bunker2023_GNz11} has \CIIIsf{} EW $=12.5\pm 1.1$\AA{}, GHZ2
\citep[$z=12.34$;][]{Castellano2024_GHZ2} has \CIIIsf{} EW $=30\pm 7$\AA{}, and GS-z12 \citep[$z=12.48$;][]{DEugenio2024_GSz12} has \CIIIsf{} EW $=30\pm 7$\AA{}. At slightly lower redshift, RXCJ2248-ID \citep[$z=6.11$][]{Topping2024_RXCJ} has \CIIIsf{} EW $=21.7\pm 0.4$\AA{}. Accompanying such strong \CIIIsf{} EW emission is evidence of high excitation and low metallicity based on rest-frame optical line ratios. From NIRSpec data alone, GN-z11, GHZ2, and GS-z12 have Ne3O2 ratios on order unity. Based on a combination of NIRSpec and MIRI spectroscopy, GHZ2 has O32$=25^{+13}_{-9}$. RXCJ2248-ID is even more extreme, with Ne3O2=$18\pm6$ and O32=$152\pm48$. According to the correlations between rest-frame optical emission-line ratios and \CIIIsf{}, and the high values of O32 and Ne3O2 for these galaxies, extreme \CIIIsf{} EWs are entirely expected. In terms of physical properties, the inferred gas-phase metallicity is $\leq 0.1$~solar and the inferred ionization parameter is $\log(U)\geq -2.0$, which are also associated with strong  \CIIIsf{} emission \citep{Tang2021,Calabro2024}. However, it is clear that a wide range of metallicity and excitation exists at $z\sim 7$, resulting in a wide range of \CIIIsf{} EW.

\section{Summary} \label{sec:summary}
We present the results of a program of Keck/MOSFIRE rest-UV spectroscopic observations for a sample of $z\sim 7$ spectroscopicially-confirmed star-forming galaxies drawn from the REBELS survey. We summarize our key results here:

\begin{enumerate}

\item Despite prior precise spectroscopic redshift constraints from ALMA \CII{} detections, we obtained no robust \CIIIsf{} detections for the 8 REBELS-MOSFIRE targets in our sample. For the targets observed during clear conditions and with both \CIIIsf{} free from sky-line contamination, these non-detections correspond to stringent upper limits of $\leq 5$\AA{} in rest-frame \CIIIsf{} EW.

\item The median \CIIIsf{} EW upper limit of our REBELS-MOSFIRE sample falls significantly below the median relation in \CIIIsf{} EW vs. redshift presented by \citet{RobertsBorsani2024}, evaluated at $z=7$.

\item The REBELS-MOSFIRE sample is significantly more massive, UV-luminous, and redder on average than the $z\sim 7$ stacked sample from  \citet{RobertsBorsani2024}. Furthermore, rest-frame optical emission-line properties for 4 of the 8 galaxies in the REBELS-MOSFIRE sample (\OIIIHb{} EW, O32, and Ne3O2) have been measured with JWST/NIRSpec subsequent to our MOSFIRE observations. These suggest both lower excitation and ionization, and higher metallicity than typical star-forming galaxies at $z\sim 7$. Given the known connection between rest-optical emission-line properties and \CIIIsf{} emission, the predicted \CIIIsf{} strengths for the REBELS-MOSFIRE sample are consistent with the upper limits we obtained.
\end{enumerate}

Although our Keck/MOSFIRE observations suggest relatively weak \CIIIsf{} emission in REBELS galaxies, we highlight that the combination of bright UV continuum luminosity and metal-richness for this sample suggests the possibility of a complementary set of rest-frame UV features that will be well within reach of NIRSpec. These include stellar wind lines from C~IV, Si~IV, and He~II, which will provide unique constraints on star-formation histories, stellar metallicities and IMF \citep[e.g.,][]{Shapley2003,Topping2020}. Furthermore, with near-IR magnitudes brighter than AB=25, it will be straightforward to detect rest-UV continuum at high signal to noise, even at medium ($R\sim 1000$) resolution. There are preliminary stacked measurements of interstellar absorption lines tracing galaxy outflows at $z>6$ \citep{Glazer2025}, but the UV-bright REBELS sample will enable such absorption-line studies of the baryon cycle for individual systems.

\section{Acknowledgments}
We acknowledge support from NSF AAG grants AST-2009313 and AST-2307622 (AES) and AST-2109066 (DPS). MT acknowledges
funding from the JWST Arizona/Steward
Postdoc in Early galaxies and Reionization (JASPER)
Scholar contract at the University of Arizona.
VG gratefully acknowledges support from ANID/CONICYT+FONDECYT Regular 1221310 and by the ANID BASAL project FB210003. MA is supported by FONDECYT grant number 1252054, and gratefully acknowledges support from ANID Basal Project FB210003 and ANID MILENIO NCN2024\_112. IDL acknowledges funding from the European Research Council (ERC) under the European Union's Horizon 2020 research and innovation program DustOrigin (ERC-2019-StG-851622), from the Belgian Science Policy Office (BELSPO) through the PRODEX project ``JWST/MIRI Science exploitation" (C4000142239) and from the Flemish Fund for Scientific Research (FWO-Vlaanderen) through the research project G0A1523N.
The authors acknowledge the Texas Advanced Computing Center (TACC) at The University of Texas at Austin for providing HPC resources that have contributed to the research results reported within this paper.
We finally wish to extend special
thanks to those of Hawaiian ancestry on whose sacred
mountain we are privileged to be guests. Without their
generous hospitality, the work presented herein would not have been possible.
%

\vspace{5mm}
\facilities{Keck(MOSFIRE)}


\software{\textsc{numpy} \citep{harris2020_numpy},
\textsc{matplotlib} \citep{Hunter2007_matplotlib},
\textsc{scipy} \citep{Virtanen2020_SciPy},
\textsc{astropy} \citep{astropy:2013, astropy:2018},
\textsc{photutils} \citep{Bradley2022_photutils},
\textsc{beagle} \citep{Chevallard2016},
\textsc{multinest} \citep{Feroz2008,Feroz2009},
\textsc{emcee} \citep{ForemanMackey2013_emcee},
\textsc{cloudy} \citep{Ferland2013},
\textsc{parsec} \citep{Bressan2012_parsec,Chen2015_parsec}
}




\bibliography{main}{}
\bibliographystyle{aasjournal}



\end{document}